\documentclass[aps,prb,twocolumn,showpacs,amsmath]{revtex4}
\usepackage{latexsym}
\usepackage{amssymb}
\usepackage{graphicx}
\usepackage{bm}
\usepackage[english]{babel}
\usepackage[latin1]{inputenc}
\usepackage{amsmath}
\usepackage{amsfonts}

\newcommand{\pdag}{{\phantom{\dagger}}}
\newcommand{\bq}{\begin{equation}}
\newcommand{\eq}{\end{equation}}
\newcommand{\bn}{\begin{eqnarray}}
\newcommand{\en}{\end{eqnarray}}

\begin{document}

\title{Ac-cotunneling through an interacting quantum dot in a magnetic field}

\author{Bing Dong}
\affiliation{Department of Physics, Shanghai Jiaotong University,
1954 Huashan Road, Shanghai 200030, China}

\author{X.L. Lei}
\affiliation{Department of Physics, Shanghai Jiaotong University,
1954 Huashan Road, Shanghai 200030, China}

\author{N. J. M. Horing}
\affiliation{Department of Physics and Engineering Physics, Stevens Institute of Technology, 
Hoboken, New Jersey 07030, USA}

\begin{abstract}

We analyze inelastic cotunneling through an interacting quantum dot subject to an ambient 
magnetic field in the weak tunneling regime under a non-adiabatic time-dependent 
bias-voltage. Our results clearly exhibit photon-assisted satellites and an overall 
suppression of differential conductance with increasing driving amplitude, which is 
consistent with experiments. We also predict a zero-anomaly in differential conductance under 
an appropriate driving frequency.       

\end{abstract}

\date{\today}

\pacs{72.40.+w, 73.23.Hk, 73.63.Kv, 03.65.Yz}

\maketitle

Recently, cotunneling\cite{Averin} through discrete levels, i.e., a quantum dot (QD), has 
attracted much attention since it determines the intrinsic limitation of accuracy of 
single-electron transistors due to leakage, and since it also involves correlation effects, 
such as the Kondo effect.\cite{cotunnelingExp} It has also been reported 
experimentally\cite{Kogan} and theoretically\cite{Ng,Goldin,Lopez,Kaminski} that external 
microwave irradiation can induce the occurrence of Kondo satellites and an overall 
suppression of the Kondo peak. However, there are few studies so far concerning 
time-dependent second-order cotunneling in the weak tunneling regime at temperatures above 
the Kondo temperature. About ten years ago, Flensberg presented an analysis for coherent 
photon-assisted cotunneling in a double-junction Coulomb blockade device in the adiabatic 
limit.\cite{Flensberg}
In this letter, we will further study the cotunneling in an interacting QD when an ac 
bias-voltage is applied between two electrodes in the non-adiabatic regime.  

We employ the s-d exchange Hamiltonian to model inelastic cotunneling through a QD in an 
ambient magnetic field, $B$, in the weak-coupling regime:
\begin{align}
H=& \,H_{0}+H_{\mathrm{I}}, & &  \label{Hamiltonian} \\
H_{0}=& \,\sum_{\eta \mathbf{k}\sigma }\varepsilon _{\eta \mathbf{k}}(t) c_{\eta 
\mathbf{k}\sigma }^{\dag }c_{\eta \mathbf{k} \sigma}^{{\phantom{\dagger}}}-\Delta_0 S^{z},\cr 
H_{\mathrm{I}}= & \,\sum_{\eta ,\eta ^{\prime },\mathbf{k}
,\mathbf{k}^{\prime }}J \bigl [\bigl(c_{\eta \mathbf{k}
\uparrow }^{\dag }c_{\eta ^{\prime }\mathbf{k}^{\prime } \uparrow}^{{\phantom{\dagger}}} - 
c_{\eta \mathbf{k}\downarrow }^{\dag }c_{\eta ^{\prime }
\mathbf{k}^{\prime }\downarrow }^{{\phantom{\dagger}}}\bigr)S^{z}\cr&
\,+c_{\eta \mathbf{k}\uparrow }^{\dag }c_{\eta ^{\prime }\mathbf{k}^{\prime} 
\downarrow}^{{\phantom{\dagger}}}S^{-} + c_{\eta \mathbf{k}\downarrow}^{\dag} 
c_{\eta^{\prime} \mathbf{k}^{\prime }\uparrow }^{{\phantom{\dagger}}} S^{+}\bigr] + H_{\rm 
dir}, \cr
H_{\rm dir}= &\, J_{\rm d} \sum_{\sigma} \bigl ( c_{L {\bf k} \sigma}^\dagger + c_{R {\bf k} 
\sigma}^\dagger \bigr ) \bigl ( c_{L {\bf k} \sigma}^\pdag + c_{R {\bf k} \sigma}^\pdag \bigr 
),  \notag
\end{align}
where $c_{\eta \mathbf{k}\sigma }^{\dagger }$ ($c_{\eta \mathbf{k}\sigma }$)
is the creation (annihilation) operator for electrons with momentum $\mathbf{k}$, 
spin-$\sigma$ in lead $\eta$ ($=\mathrm{L,R}$). The energies $\varepsilon _{\eta 
\mathbf{k}}(t)= \varepsilon _{\eta \mathbf{k}}^0+eV_{\eta}(t)$ include a rigid shift of the 
Fermi energy of the electrons in the leads due to the applied time-dependent bias-voltage 
$V_{\eta}(t)=V_{\eta}^0 + v_{\eta}\cos(\Omega t)$ with $V_{\eta}^0$ ($v_{\eta}$) being the 
amplitude of the dc(ac) part of the bias-voltage. Here, we assume that the Fermi energies of 
two leads are zero at equilibrium and $V_{L}^0=-V_{R}^0=eV_0/2$.   
$\Delta_0=g_e\mu _{B}B$ is the static magnetic-field $B$-induced Zeeman energy. 
${\bf S}\equiv (S^x, S^y, S^z)$ are Pauli spin operators of electrons in the QD 
[$S^{\pm}\equiv S^x \pm i S^y$], and $J$ is the exchange coupling constant. 
$H_{\rm dir}$ is the potential scattering term with $2J_{\rm d}=J$.      
As in our previous paper,\cite{Dong} we can rewrite the tunneling term, $H_{\rm I}$ in 
Eq.~(\ref{Hamiltonian}), as a sum of three products of two variables: 
\bq
H_{\rm I}=Q^z S^z+ Q^+ S^- + Q^- S^+ + Q^{\hat 1},
\eq
with the generalized coordinates $Q^{z(\pm)}$ of reservoir variables as 
\begin{align}
Q^{z}=&\, \sum_{\eta,\eta'} Q_{\eta\eta'}^{z}= \sum_{\eta,\eta', {\bf k},{\bf k}'} J \bigl( 
c_{\eta {\bf k} \uparrow}^\dag c_{\eta' {\bf k}' \uparrow}^\pdag - c_{\eta {\bf k} 
\downarrow}^\dag c_{\eta' {\bf k}' \downarrow}^\pdag \bigr), \label{Qz} \\
Q^{+}=&\, \sum_{\eta,\eta'} Q_{\eta\eta'}^{+}= \sum_{\eta, \eta', {\bf k},{\bf k}'} J c_{\eta 
{\bf k} \uparrow}^\dag c_{\eta' {\bf k}' \downarrow}^\pdag, \label{Q+}\\
Q^{-}=&\, \sum_{\eta,\eta'} Q_{\eta\eta'}^{-}= \sum_{\eta, \eta', {\bf k},{\bf k}'} J c_{\eta 
{\bf k} \downarrow}^\dag c_{\eta' {\bf k}' \uparrow}^\pdag, \label{Q-}
\end{align}
and $Q^{\hat 1}=H_{\rm dir}$.
In the following, we will use units where $\hbar=k_{B}=e=1$.

As in our previous studies of inelastic cotunneling through an interacting QD in the weak 
tunneling limit, we employ a generic quantum Langevin equation 
approach\cite{Ackerhalt,Cohen,Smirnov} to establish a set of quantum Bloch equations for the 
description of the dynamics of a single spin [modeled by Eq.~(\ref{Hamiltonian})] explicitly 
in terms of the response and correlation functions of free reservoir variables. This 
procedure provides explicit analytical expressions for the nonequilibrium magnetization and 
cotunneling current for arbitrary dc bias-voltage and temperature.\cite{Dong,Dong1} Here, we 
generalize our previous derivations to the time-dependent case in the non-adiabatic and 
high-frequency regime. 

In the derivation, we proceed with the Heisenberg equation of motion for the Pauli spin 
operators and the lead operators, and then formally integrate these equations from the 
initial time $0$ to $t$ exactly to all orders of $J$. Next, under the assumption that the 
time scale of decay processes is much slower than that of free evolution, we replace the 
time-dependent
operators involved in the integrals of these EOM's approximately in terms of their free 
evolutions. Thirdly, these EOM's are expanded in powers of $J$ up to second order, resulting 
in non-Markovian dynamic equations for the time evolution of the QD spin variables in a 
compact form:
\bn
\dot S^{z}(t)&=& -2 \int_{-\infty}^t dt' \left ( e^{-i\Delta_0 \tau} + e^{i\Delta_0 \tau} 
\right ) C(t,t') S^{z}(t') \cr
&& - \int_{-\infty}^t dt' \left ( e^{-i\Delta_0 \tau} - e^{i\Delta_0 \tau} \right ) R(t,t'), 
\label{eom:sz} \\
\dot S^{\pm}(t)&=& \mp i \Delta_0\, S^\pm(t) - 2 \int_{-\infty}^t dt' C(t,t') S^{\pm}(t') \cr
&& - 2 \int_{-\infty}^t dt' e^{\mp i\Delta_0 \tau} C(t,t') S^{\pm}(t'), \label{eom:spm}
\en
with $\tau=t-t'$. The correlation function, $C(t,t^{\prime })$, and 
the response function, $R(t,t^{\prime })$,
of free reservoir variables (tagged by subscript ``$o$") are defined as:
\bq
C(R)(t,t^{\prime }) ={\frac{1}{2}}\theta (\tau )\langle [ Q^{\pm}_{o}(t), 
Q^{\mp}_{o}(t^{\prime })]_{+(-)}\rangle.  \label{randcf}
\eq

Special attention must be paid to the free reservoir variables due to the time-dependent 
energies $\varepsilon _{\eta \mathbf{k}}(t)$:
\bq
c_{\eta {\bf k} \sigma}^o(t) = e^{-i\int_{t'}^t \varepsilon_{\eta {\bf k}}(\tau) d\tau} 
c_{\eta {\bf k} \sigma}(t').
\eq
Therefore, the kernels $C(R)(t,t')$ become {\em double-time-dependent} functions due to the 
lack of time-translation-invariance stemming from the ac-bias:
\bn
C(R)(t,t') &=& {1\over 2}\theta(\tau) \langle [Q_{o}^{+}(t), Q_{o}^{-}(t')]_\pm \rangle \cr
&=& {1\over 2} \theta(\tau) J^2 \sum_{\eta,\eta',\xi,\xi'} \sum_{{\bf k},{\bf k}',{\bf 
q},{\bf q}'} \langle [c_{\eta{\bf k} \uparrow}^\dagger (t) c_{\eta' {\bf k}' 
\downarrow}^\pdag (t), \cr
&& c_{\xi {\bf q} \downarrow}^\dagger(t') c_{\xi' {\bf q}' \uparrow}^\pdag (t')]_\pm \rangle 
\cr
&=& {1\over 2} \theta(\tau) J^2 \cr
&& \times \sum_{\eta,\eta',\xi,\xi'} \sum_{{\bf k},{\bf k}',{\bf q},{\bf q}'} e^{i\int_{t'}^t 
d\tau [\epsilon_{\xi'{\bf q}'}(\tau)- \epsilon_{\xi {\bf q}}(\tau)] } \cr
&& \times \left [ \langle c_{\eta {\bf k} \uparrow}^\dagger (t) c_{\xi' {\bf q}' 
\uparrow}^\pdag (t) \rangle \langle c_{\eta' {\bf k}' \downarrow}^\pdag (t) c_{\xi {\bf q} 
\downarrow}^\dagger (t) \rangle  \right. \cr
&& \left. \pm \langle c_{\xi {\bf q} \downarrow}^\dagger (t) c_{\eta' {\bf k}' 
\downarrow}^\pdag (t) \rangle \langle c_{\xi' {\bf q}' \uparrow}^\pdag (t) c_{\eta {\bf k} 
\uparrow}^\dagger (t) \rangle \right ]\cr
&=& \frac{1}{2}\theta(\tau) \sum_{\eta} g \int d\epsilon d\epsilon' e^{i(\epsilon-\epsilon') 
\tau} \cr
&& \times \left \{ f_{\eta}(\epsilon) \left [ 1-f_{\eta}(\epsilon') \right ] \pm 
f_{\eta}(\epsilon') \left [ 1-f_{\eta}(\epsilon) \right ] \right \} \cr
&& + \frac{1}{2}\theta(\tau) g \int d\epsilon d\epsilon' \left [ 
e^{i(\epsilon-\epsilon')\tau} \right. \cr
&& \times e^{i\int_{t'}^t d\tau' V_{\rm ac} \cos(\Omega \tau')} \pm  
e^{-i(\epsilon-\epsilon')\tau} \cr
&& \left. \times e^{-i\int_{t'}^t d\tau' V_{\rm ac} \cos(\Omega \tau')} \right ] 
\left \{ f_{L}(\epsilon) \left [ 1-f_{R}(\epsilon') \right ] \right. \cr
&& \left. \pm f_{R}(\epsilon') \left [ 1-f_{L}(\epsilon) \right ] \right \},
\en
with $g\equiv J^{2} \rho_{0}^{2}$, $V_{\rm ac}=v_L-v_R$, and the Fermi-distribution function 
is $f_{\eta}(\epsilon)=\left [1+ e^{(\epsilon-\mu_{\eta})/T} \right ]^{-1}$ ($T$ is the 
temperature). Here we assume the two electrodes to be Markov-type reservoirs with a constant 
density of states $\rho_0$. The kernels reduce exactly to our previous results, Eq.~(B8) in 
Ref.~\onlinecite{Dong}, if there is no ac-bias or with the same ac amplitude in the left and 
right leads. 

In the presence of a periodic ac-bias, the spin variables $S^{z(\pm)}(t)$ naturally depend 
periodically on $t$ with a period ${\cal T}_{\rm ac}=2\pi/\Omega$. As a result, the full 
solutions of Eqs.~(\ref{eom:sz}) and (\ref{eom:spm}) can be formally written as a 
superposition of all harmonics
\bq
S^{z(\pm)}(t)=\sum_{n=-\infty}^{\infty} S_{(n)}^{z(\pm)} e^{-i n \Omega t}.
\eq
Employing this expansion in the dynamic equations (\ref{eom:sz}) and (\ref{eom:spm}), an 
infinite set of linear equations results in which the $S_{(n)}^{z(\pm)}$ are coupled with 
each other via the kernels. To obtain a solution for the spin variables, one has to terminate 
this infinite chain at a chosen order and then solve the resulting equations in a recursive 
way. However, in the limit of high frequencies, $\Omega\gg \Gamma$ (tunneling rate) and $T$ 
(temperature), of interest in this letter, the ac-bias oscillates so fast that an electron 
experiences many cycles of the ac-bias during its presence inside the dot, and thus can not 
sense the details of the dynamics within one period ${\cal T}_{\rm ac}$. In this 
non-adiabatic limit, one can approximately replace the kernels by a time-average with respect 
to the center-of-mass of time ${\bar \tau}=t+t'$:\cite{Tien,Vicari}
\bq
C(R)(t,t')\approx C(R)^{(0)}(\tau) = \frac{1}{{\cal T}_{\rm ac}} \int_0^{{\cal T}_{\rm ac}} d 
{\bar \tau} C(R)(\tau,\bar \tau).
\eq
By the same token, one can retain only the stationary part of the spin variables and neglect 
the rapidly oscillatory parts, leading to detailed balance equations in a Markov 
approximation by making the replacement $\int_{-\infty}^t d\tau\Rightarrow 
\int_{-\infty}^\infty d\tau$ in Eqs.~(\ref{eom:sz}) and (\ref{eom:spm}):
\bn
0&=& -2 \left (C_\omega^{(0)}(-\Delta_0) + C_\omega^{(0)}(\Delta_0)\right ) S_{(0)}^{z} \cr
&& + R_\omega^{(0)}(\Delta_0) - R_\omega^{(0)}(-\Delta_0),  \label{fun:sz} \\
0 &=& \mp i \Delta_0\, S_{(0)}^\pm - 2 \left [ C_\omega^{(0)}(0) + C_\omega^{(0)}(\mp 
\Delta_0) \right ] S_{(0)}^{\pm}. 
\en
Here, $C(R)_\omega^{(0)}(\omega)$ are the Fourier transforms of the time-averaged kernels 
$C(R)^{(0)}(\tau)$:     
\bn
C_\omega^{(0)}(\omega)&=& \pi g T \varphi \left ( \frac{\omega}{T} \right ) + \frac{\pi}{2} g 
T \sum_{n=-\infty}^{\infty} J_n^2\left ( \frac{V_{\rm ac}}{\Omega} \right ) \cr
&& \times \left [ \varphi \left ( \frac{\omega + V+ n\Omega}{T}\right ) + \varphi \left ( 
\frac{\omega- V-n\Omega}{T}\right ) \right ], \cr
&& \\
R_\omega^{(0)}(\omega)&=& \pi g \left [ 1 + \sum _{n=-\infty}^{\infty} J_n^2\left ( 
\frac{V_{\rm ac}}{\Omega} \right ) \right ] \omega, \label{response}
\en
with $\varphi(x)\equiv x \coth  ( x/2 )$ and $J_n(x)$ is the Bessel function of order $n$. To 
derive these equations (\ref{fun:sz})-(\ref{response}), we use the relation
\bq
e^{ix \sin (\omega t)}=\sum_{n=-\infty}^{\infty} J_{n}(x) e^{in \omega t}.
\eq

The solution of Eq.~(\ref{fun:sz}) yields the nonequilibrium magnetization of the QD subject 
to an ac-bias voltage as
\bq
S_{(0)}^z = \frac{R_{\omega}^{(0)}(\Delta_0)}{2C_{\omega}^{(0)}(\Delta_0)}. 
\label{magnetization}
\eq
This formula is our central result, which can be regarded as a direct generalization of the 
dc nonequilibrium magnetization\cite{Dong,Parcollet,Paaske2} of a QD under a non-adiabatic 
high-frequency field. Obviously, it reduces exactly to previous results in absence of 
ac-bias, $V_{\rm ac}=0$.\cite{Dong,Parcollet,Paaske2} As an illustration, we exhibit in 
Fig.~1(a) the dependence of the magnetization, $S_{(0)}^z$, on dc bias-voltage for the 
driving frequency $\Omega/\Delta_0=0.5$. It should be noted that $S_{(0)}^z$ exhibits 
different behaviors with increasing ac-amplitude $V_{\rm ac}$. For small dc bias voltage, the 
QD spin is fully polarized due to the nonzero external magnetic field, and it is gradually 
quenched with increasing dc bias voltage. Application of an ac bias tends to quench the spin 
polarization more rapidly. This tendency suggests that the ac bias plays a role in dephasing 
the electronic tunneling processes in a QD, as analyzed in Ref.~\onlinecite{Kaminski}.   

\begin{figure}[htb]
\includegraphics [width=8.5cm,height=4cm,angle=0,clip=on]{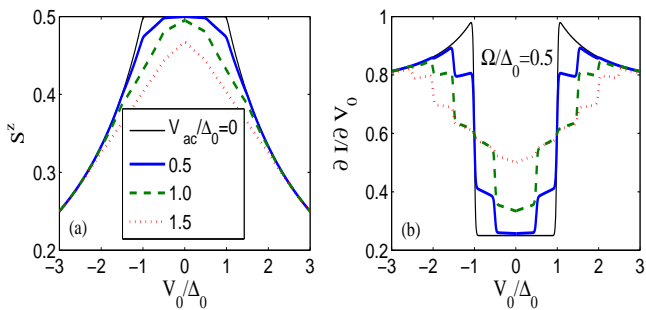}
\caption{Nonequilibrium magnetization, $S_{(0)}^z$ (a); and the differential conductance, 
$dI/dV_0$ (in units of $2ge^2/h$) (b); as functions of dc bias-voltage, $V_0$, for various 
amplitudes of the ac-bias with a fixed driving frequency $\Omega/\Delta_0=0.5$. The 
temperature we use in the calculation is $T/\Delta_0=0.01$.} \label{fig1}
\end{figure}

Proceeding to the calculation of tunneling current, the current operator through the QD is 
defined as the time rate of change of charge density $N_{\eta}=\sum_{{\bf k},\sigma} 
c_{\eta{\bf k}\sigma}^\dagger c_{\eta{\bf k} \sigma}^\pdag$ in lead $\eta$: $I_{\eta}(t) = 
\dot N_{\eta }$. From linear-response theory we have
\bq
I(t) = \langle I_{\eta }(t) \rangle =-i\int_{-\infty}^t dt' \langle [ J_{\eta }(t), H_{\rm 
I}(t')]_- \rangle. \label{currentsigma}
\eq
Because the dc component of current is easily measurable experimentally, we compute the 
time-averaged current $I = \frac{1}{{\cal T}_{\rm ac}} \int_0^{{\cal T}_{\rm ac}} dt I(t)$. 
Performing the same high-frequency and Markov approximations as above, we obtain the 
time-averaged currents:
\bn
I &=& 4\pi g \sum_{n=-\infty}^{\infty} J_n^2\left ( \frac{V_{\rm ac}}{\Omega} \right ) ( V + 
n\Omega) + 2 \pi g T S_{(0)}^z \cr
&& \times \sum_{n=-\infty}^{\infty} J_n^2\left ( \frac{V_{\rm ac}}{\Omega} \right ) 
\left [ \varphi \left ( {\Delta_0 -V -n\Omega\over T} \right ) \right. \cr
&& \left. - \varphi \left ( {\Delta_0 +V +n\Omega\over T} \right ) \right ], \label{ic}
\en
which is a generalization of Tien-Gordon-type formula in cotunneling current.\cite{Tien} It 
should be noted that our expression for photon-assisted cotunneling current is valid in the 
high frequency limit, whereas Flensberg derived a perturbative current under pump conditions 
in the low-frequency limit.\cite{Flensberg}   

\begin{figure}[htb]
\includegraphics [width=8.5cm,height=4cm,angle=0,clip=on]{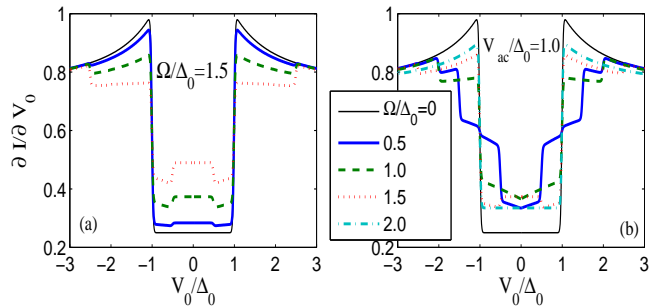}
\caption{Calculated differential conductance vs. dc-bias. (a) results for a relatively high 
driving frequency $\Omega/\Delta_0=1.5$ for various ac-amplitudes, as in Fig.~1; (b) the 
ac-frequency dependence of $dI/dV_0$ for a fixed driving amplitude $V_{\rm ac}/\Delta_0=1.0$. 
Other parameters are the same as in Fig.~1.} \label{fig2}
\end{figure}

We plot the dc bias-voltage-dependent differential conductance, $dI/dV_0$, in Figs.~1(b) and 
2. Obviously, the differential conductance shows some satellites at $V_0=\pm 
(\Delta_0-n\Omega)$ superimposed on the characteristic jump at $V_0=\pm \Delta_0$ in the 
presence of an ac-bias. These satellites arise physically from photon-assisted spin-flip 
cotunneling, i.e., albeit $|V_0|<\Delta_0$, the spin-flip cotunneling process can still 
become energetically activated by an electron absorbing photon quanta to compensate for the 
energy difference. Moreover, an overall suppression is observed with increasing ac-amplitude, 
which is qualitatively consistent with the experimental results.\cite{Kogan} More 
interestingly, we find that $dI/dV_0$ exhibits a transition from peak-splitting to 
zero-bias-anomaly if the driving frequency is higher than $\Delta_0$. This behavior can be 
ascribed to the fact that a driving field with an appropriately high frequency can spur 
spin-flip cotunneling, notwithstanding $|V_0|<(\Omega-\Delta_0)$; in contrast, when the 
dc-bias increases to $\Delta_0>|V_0|>(\Omega-\Delta_0)$, spin-flip events become inactive 
instead. The $dI/dV_0$ curve recovers peak-splitting behavior if $\Omega \geq 2\Delta_0$, as 
shown in Fig.~2(b).   

In summary, we have generalized the generic Langevin equation approach to study 
photon-assisted cotunneling through an interacting QD in the non-adiabatic and high frequency 
regime, deriving explicit analytic expressions for the dc components of nonequilibrium 
magnetization and current with a generalized Tien-Gordon-type form. Our results show that 
applying an ac-bias is an important method for tuning the $I$-$V$ characteristics. 
Considering experiments\cite{cotunnelingExp} in which a static magnetic field $B=11$ T is 
applied, $\Delta_0\simeq 0.1$ meV, and the ac-frequency is $\Omega\sim 12-50$ GHz with 
ac-amplitude $V_{\rm ac}\leq 0.15$ mV, all the parameters are easily accessible 
experimentally and they satisfy the non-adiabatic condition $\Omega\gg \Gamma$ ($\sim 30$ 
$\mu$eV).

This work was supported by Projects of the National Science Foundation of China, the Shanghai 
Municipal Commission of Science and Technology, the Shanghai Pujiang Program, and Program for 
New Century Excellent Talents in University (NCET).

\end{document}